\documentclass[twocolumn,showpacs,preprintnumbers,amsmath,amssymb, showkeys]{revtex4}

\usepackage{graphicx}
\usepackage{dcolumn}
\usepackage{bm}
\usepackage{epsf}

\begin{document}


\title{Time evolution of an entangled initial state in coupled quantum dots with Coulomb correlations}

\author{N.\,S.\,Maslova}%
 \email{spm@spmlab.phys.msu.ru}
 \author{V.\,N.\,Mantsevich}
 \altaffiliation{vmantsev@spmlab.phys.msu.ru}
\author{P.\,I.\,Arseyev}
 \altaffiliation{ars@lpi.ru}
\affiliation{Moscow State University, Department of  Physics, 119991
Moscow, Russia\\~\\ P.N. Lebedev Physical institute of RAS, 119991,
Moscow, Russia}

\date{\today }
9 pages, 6 figures
\begin{abstract}
We analyzed the dynamics of the initial singlet electronic state in
the two interacting single-level quantum dots (QDs) with Coulomb
correlations, weakly tunnel coupled to an electronic reservoir. We
obtained correlation functions of all orders for the electrons in
the QDs by decoupling high-order correlations between localized and
band electrons in the reservoir. We proved that for arbitrary mixed
state the concurrence and entanglement can be determined from the
average value of particular combinations of electron's pair
correlation functions. Analysis of the pair correlation functions
time evolution allows to follow the changes of concurrence and
entanglement during the relaxation processes. We investigated the
dependence of concurrence on the value of Coulomb interaction and
the energy levels spacing and found it's non-monotonic behavior in
the non-resonant case. We also demonstrated that the behavior of
pair correlation functions for two-electron entangled state in
coupled QDs points to the fulfillment of the Hund's rule for the
strong Coulomb interaction. We revealed the appearance of dynamical
inverse occupation of the QDs energy levels during the relaxation
processes. Our results open up further perspectives in solid state
quantum information based on the controllable dynamics of the
entangled electronic states.
\end{abstract}

\pacs{73.20.Hb, 73.23.Hk, 73.40.Gk}
\keywords{D. Entangled electronic states; D. Coulomb correlations; D. Quantum dots; D. Relaxation}
\maketitle

\section{Introduction}

Now a days, low-dimensional semiconductor structures with a small
number of electrons attract much attention. The main reason for that
interest is the progress in technological procedure that allows to
fabricate nanostructures with high precision
\cite{Jacak98},\cite{Wiel02}. In recent years experimental technique
gives possibility to create vertically aligned strongly interacting
QDs with only one of them coupled to the continuous spectrum states
\cite{Vamivakas},\cite{Stinaff}. This side-coupled geometry gives an
opportunity to fabricate many-particle states with various charge
and spin configurations in the small size structures
\cite{Kikoin0},\cite{Arseyev_2},\cite{Mantsevich},\cite{Shulman},\cite{Mantsevich_1}.
Considerable progress was achieved in fabrication of lateral QDs,
which are extremely tunable by means of individual electrical gates
\cite{Kastner_1},\cite{Ashoori}. This advantage reveals in the
possibility of single electron localization in the system of several
coupled QDs \cite{Chan} and charge states manipulations in the
artificial molecules \cite{Arseyev_1}. There are a lot of possible
applications of this field in nanoelecttronics \cite{Dowling},
including quantum information processing \cite{Nielsen}. Double QDs
play an important role in the problem of quantum information
processing \cite{Loss},\cite{Bennett}. Most of the proposed schemes
for quantum computation deal with the spin control because of the
long decoherence times \cite{Hanson}. But now due to the development
of light sources the control on electric charge in low-dimensional
systems is produced both by gate voltages
\cite{Murgida},\cite{Kataoka} and laser pulses
\cite{Putaja},\cite{Saelen}. Creating, controlling and detecting
entangled states in ultra small condensed matter systems is one of
the most important problems for future quantum computation
applications \cite{Engel} and for the development of new electronic
devices based on semiconductor nanostructures. It is important to be
able to prepare interacting few-level systems with different initial
states
\cite{Bayer},\cite{Arseyev},\cite{Creatore},\cite{Tsukanov},\cite{Yokoshi}
- from simple product states to complex entanglements. Various ideas
for entangling of spatially separated electrons were proposed, such
as, by splitting Cooper pairs \cite{Burkard} or by spin manipulation
in QDs \cite{Sanchez},\cite{Ciccarello}. In double QDs an entangled
state can be obtained by putting the electrons into a singlet ground
state \cite{Loss},\cite{Burkard_1},\cite{Blaauboer}. Electron
transport in coupled QDs is governed by Coulomb interaction between
localized electrons, by the ratios between tunneling transfer
amplitudes and the quantum dots coupling and of course by the
initial conditions \cite{Maslova},\cite{Maslova_1}. To integrate
quantum dots in a small quantum circuits it is necessary to analyze
the influence of non-equilibrium charge distribution, relaxation
processes and non-stationary effects on the electron transport
through the system. So the problem of charge relaxation due to the
tunneling processes between QDs coupled to an electronic reservoir
in the presence of Coulomb interaction is really vital.
Consequently, the detailed analysis of time evolution of initial
singlet entangled two-electron state in the system of interacting
DQs with Coulomb correlations is an important problem, which may
have further implications for quantum information in nanoscale
devices.

In this paper we consider charge relaxation in the double QD due to
the coupling to an electronic reservoir. Tunneling from the first QD
to the continuum is possible only through the second dot. We
obtained the closed system of equations for time evolution of the
localized electrons filling numbers and pair correlation functions
which exactly takes into account all order correlation for localized
electrons. We decoupled the high order correlation functions between
conduction electrons in the reservoir and electrons localized in the
QDs. In such an approximation the electrons distribution in the
reservoir is not influenced by changing of an electronic states in
the coupled QDs. For QDs weakly coupled to the reservoir the
proposed decoupling scheme is a good approximation. We considered
system relaxation from initial singlet entangled two-electron state
and took into account Coulomb correlations within both QDs. Such
state can be prepared experimentally as a ground two-particle state
for vertically aligned strongly coupled QDs if interaction with
substrate is extremely weak in comparison with interdot coupling and
interaction with the other lead (reservoir). The interaction with
reservoir (for example STM tip) is switched on at the initial time
moment. We proved that for a mixed state the concurrence and
entanglement can be expressed through the average value of
particular combinations of electron's pair correlation functions. We
performed the analysis of the pair correlation functions time
evolution, which allowed us to follow the changes of concurrence and
entanglement during the relaxation process. We found some
peculiarities in electrons filling numbers dynamics arising due to
the Coulomb correlation effects. We demonstrated the appearance of
dynamical inverse occupation in the proposed system and revealed the
fulfilment of Hund's rule \cite{H1},\cite{H2},\cite{H3}.

\section{Model}
We consider a system of coupled QDs with the energy levels
$\varepsilon_1$ and $\varepsilon_2$ connected to an electronic
reservoir (see Fig.\ref{Fig.1}). At the initial time moment the
interaction between the QD with energy level $\varepsilon_2$ (second
QD) and electronic reservoir ($\varepsilon_p$) is switched on. In
the absence of interaction with the reservoir two-electronic states
in the coupled QDs are entangled in the presence of Coulomb
correlations. We model the system by the Hamiltonian $\hat{H}$:

\begin{figure}[h]
\includegraphics[width=60mm]{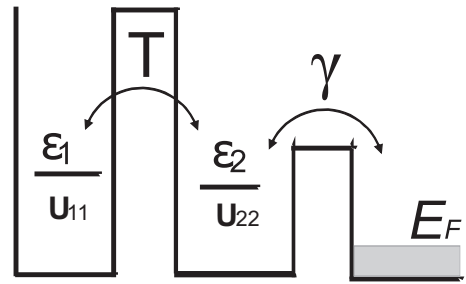}%
\caption{Scheme of the proposed model. The system of interacting QDs
is coupled to an electronic reservoir by means of tunneling rate
$\gamma=\pi\nu_0t^{2}$.} \label{Fig.1}
\end{figure}

\begin{eqnarray}
\hat{H}=\hat{H}_{D}+\hat{H}_{tun}+\hat{H}_{res}.
\end{eqnarray}

The Hamiltonian $\hat{H_{D}}$ of interacting QDs

\begin{eqnarray}
\hat{H}_{D}&=&\sum_{l=1,2\sigma}\varepsilon_{l}c^{+}_{l\sigma}c_{l\sigma}+\sum_{l=1,2\sigma}U_{l}n_{ll\sigma}n_{ll-\sigma}+\nonumber\\&+&\sum_{\sigma}T(c_{1\sigma}^{+}c_{2\sigma}+c_{2\sigma}^{+}c_{1\sigma}),
\end{eqnarray}

contains the spin-degenerate levels $\varepsilon_l$ (indexes $l=1$
and $l=2$ correspond to the first and to the second QD) and the
on-site Coulomb energy $U_{l}$ for double occupation of the dots.
The creation/annihilation of an electron with spin $\sigma=\pm1/2$
within the dot is denoted by $c^{+}_{l\sigma}/c_{l\sigma}$ and
$n_{ll\sigma}$ is the corresponding filling number operator. The
coupling between the dots is described by the tunneling transfer
amplitude $T$ which is considered to be independent of momentum and
spin.

The reservoir is modeled by the Hamiltonian:

\begin{eqnarray}
\hat{H}_{res}=\sum_{p\sigma}\varepsilon_{p}c^{+}_{p\sigma}c_{p\sigma},
\end{eqnarray}

in which $c^{+}_{p\sigma}/c_{p\sigma}$ creates/annihilates an
electron with spin $\sigma$ and momentum $p$ in the lead. The
coupling between the second dot and the reservoir is described by
the Hamiltonian:

\begin{eqnarray}
\hat{H}_{tun}=\sum_{p\sigma}t(c_{p\sigma}^{+}c_{2\sigma}+c_{p\sigma}^{+}c_{2\sigma}),
\end{eqnarray}

where $t$ is the tunneling amplitude, which is considered to be
independent on momentum and spin. Taking into account a constant
density of states in the reservoir $\nu_0$, the tunnel coupling rate
$\gamma$ is defined as $\gamma=\pi\nu_0t^{2}$. Interaction between
the second QD and the reservoir is switched on at the initial time
moment.

In the absence of coupling with the reservoir correlated two
particle pure states for electrons with opposite spins are described
by the wave function

\begin{eqnarray}
|\psi\rangle=\alpha|\uparrow\downarrow\rangle|0\rangle+\beta|\downarrow\rangle|\uparrow\rangle+
\gamma|\uparrow\rangle|\downarrow\rangle+\delta|0\rangle|\uparrow\downarrow\rangle.
\label{wave}\end{eqnarray}

For non-zero value of Coulomb interaction in the system this state
is entangled. Such state with the lowest emergy can be
experimentally prepared in strongly coupled vertically aligned QDs
in the case when interaction with substrate is extremely weak in
comparison with interdot coupling and interaction with the other
lead (reservoir, for example STM tip).

We set $\hbar=1$ and derive the kinetic equations for bilinear
combinations of Heisenberg operators $c_{l\sigma}^{+}/c_{l\sigma}$

\begin{eqnarray}
c_{1\sigma}^{+}c_{1\sigma}=\hat n_{11}^{\sigma}(t);\quad
c_{2\sigma}^{+}c_{2\sigma}=\hat n_{22}^{\sigma}(t);\nonumber\\
c_{1\sigma}^{+}c_{2\sigma}=\hat n_{12}^{\sigma}(t);\quad
c_{2\sigma}^{+}c_{1\sigma}=\hat n_{21}^{\sigma}(t).
\end{eqnarray}
Kinetic equations describe time evolution of the electron filling
numbers in each QD

\begin{eqnarray}
i\frac{\partial}{\partial
t}\hat n_{11}^{\sigma}&=&-T(\hat n_{21}^{\sigma}-\hat n_{12}^{\sigma}),\nonumber\\
i\frac{\partial}{\partial t}\hat n_{22}^{\sigma}&=&T(\hat
n_{21}^{\sigma}-\hat n_{12}^{\sigma})- 2i\gamma \hat
n_{22}^{\sigma},\nonumber\\
i\frac{\partial}{\partial t}\hat n_{21}^{\sigma}&=&T(\hat
n_{22}^{\sigma}-\hat n_{11}^{\sigma}) +[\xi+U_{1}\hat
n_{11}^{-\sigma}] \hat n_{21}^{\sigma}-\nonumber\\&-& U_{2}\hat
n_{21}^{\sigma}\hat n_{22}^{-\sigma}
-i\gamma \hat n_{21}^{\sigma},\nonumber\\
i\frac{\partial}{\partial t}\hat n_{12}^{\sigma}&=&-T(\hat
n_{22}^{\sigma}-\hat n_{11}^{\sigma})-[\xi+U_{1}\hat
n_{11}^{-\sigma}] \hat n_{12}^{\sigma} +\nonumber\\&+& U_{2}\hat
n_{12}^{\sigma}\hat n_{22}^{-\sigma}-i\gamma \hat n_{12}^{\sigma},
\label{system}
\end{eqnarray}

where $\xi=\varepsilon_1-\varepsilon_2$ is the detuning between the
energy levels in the QDs. The system of Eqs. (\ref{system}) contains
expressions for the pair correlators $\hat{n}_{1}^{-\sigma}\hat
n_{21}^{\sigma}$ and $\hat n_{1}^{-\sigma}\hat n_{12}^{\sigma}$,
which also determine relaxation and consequently have to be
calculated. We'll neglect high order correlation functions between
localized and reservoir electrons and fulfill averaging over
electron states in the reservoir.

Let us introduce the following designation for the pair correlations
operators $\widehat{K}^{\sigma\sigma^{'}}_{lrl^{'}r^{'}}$ and their
averaged values
$K^{\sigma\sigma^{'}}_{lrl^{'}r^{'}}=<c_{l\sigma}^{+}c_{r\sigma}c_{l^{'}\sigma^{'}}^{+}c_{r^{'}\sigma^{'}}>$.
We'll consider only the paramagnetic case $<\hat
n_{l}^{\sigma}>=<\hat n_{l}^{-\sigma}>$. The system of equations for
pair correlators can be written in the compact matrix form (symbol
$[\quad]$ means commutation and symbol $\{\quad\}$-
anti-commutation)

\begin{eqnarray}
i\frac{\partial}{\partial
t}\widehat{\textbf{K}}=[\widehat{\textbf{K}},\widehat{H}^{'}]+\{\widehat{\textbf{K}},\widehat{\Gamma}\}+\widehat{\Upsilon},
\label{system_compact}
\end{eqnarray}

where $\widehat{\textbf{K}}$ is the pair correlators matrix

\begin{eqnarray}
\widehat{\textbf{K}}= \left(\begin{array}{ccccc}
K_{2211}^{\sigma-\sigma} & K_{1211}^{\sigma-\sigma} & K_{2221}^{\sigma-\sigma} & K_{1221}^{\sigma-\sigma}\\
K_{2111}^{\sigma-\sigma} & K_{1111}^{\sigma-\sigma} & K_{2121}^{\sigma-\sigma} & K_{1121}^{\sigma-\sigma}\\
K_{2212}^{\sigma-\sigma} & K_{1212}^{\sigma-\sigma} & K_{2222}^{\sigma-\sigma} & K_{1222}^{\sigma-\sigma}\\
K_{2112}^{\sigma-\sigma} & K_{1112}^{\sigma-\sigma} & K_{2122}^{\sigma-\sigma} & K_{1122}^{\sigma-\sigma}\\
\end{array}\right),
\label{correlators}
\end{eqnarray}

matrix $\widehat{H}^{'}$ has the following form

\begin{eqnarray}
\widehat{H}^{'}= \left(\begin{array}{ccccc}
0 & T & T & 0\\
T & \xi+U_{1} & 0 & T\\
T & 0 & -\xi+U_{2} & T\\
0 & T & T & 0\\
\end{array}\right)
\end{eqnarray}

and the relaxation matrix $\widehat{\Gamma}$ is denoted as

\begin{eqnarray}
\widehat{\Gamma}= \left(\begin{array}{ccccc}
-i\gamma & 0 & 0 & 0\\
0 & 0 & 0 & 0\\
0 & 0 & -2i\gamma & 0\\
0 & 0 & 0 & -i\gamma\\
\end{array}\right).
\end{eqnarray}

It is clearly evident that Eqs. (\ref{system_compact}) contain
expressions for the high-order correlators
$K_{121122}^{\sigma-\sigma-\sigma}$ and
$K_{211122}^{\sigma-\sigma-\sigma}$. Their contribution can be
written in the matrix form $\widehat{\Upsilon}$

\begin{eqnarray}
\left(\begin{array}{ccccc}
0 & U_{2}K_{121122}^{\sigma-\sigma-\sigma} & U_{1}K_{211122}^{\sigma-\sigma-\sigma} & 0\\
-U_{2}K_{211122}^{\sigma-\sigma-\sigma} & 0 & 0 & -U_{2}K_{211122}^{\sigma-\sigma-\sigma}\\
-U_{1}K_{121122}^{\sigma-\sigma-\sigma} & 0 & 0 & -U_{1}K_{121122}^{\sigma-\sigma-\sigma}\\
0 & U_{2}K_{121122}^{\sigma-\sigma-\sigma} & U_{1}K_{211122}^{\sigma-\sigma-\sigma} & 0\\
\end{array}\right).
\end{eqnarray}

The system of equations (\ref{system}),(\ref{system_compact}) for
the two electronic pure entangled state $|\psi\rangle$ time
evolution in the coupled QDs connected with the reservoir can be
solved with the following initial conditions:
$n_{11}^{\sigma}(0)=\alpha^{2}+\beta^{2}$;
$n_{12}^{\sigma}(0)=n_{21}^{\sigma}(0)=\alpha\gamma+\beta\delta$;
$n_{22}^{\sigma}(0)=\delta^{2}+\gamma^{2}$;
$K_{1111}^{\sigma-\sigma}=\alpha^{2}$;
$K_{2222}^{\sigma-\sigma}=\delta^{2}$;
$K_{1122}^{\sigma-\sigma}=\beta^{2}$;
$K_{2211}^{\sigma-\sigma}=\gamma^{2}$;
$K_{1221}^{\sigma-\sigma}=K_{2112}^{\sigma-\sigma}=\beta\gamma$;
$K_{2121}^{\sigma-\sigma}=K_{1212}^{\sigma-\sigma}=\alpha\delta$;
$K_{1211}^{\sigma-\sigma}=K_{2111}^{\sigma-\sigma}=\gamma\alpha$;
$K_{1112}^{\sigma-\sigma}=K_{1121}^{\sigma-\sigma}=\alpha\beta$;
$K_{1222}^{\sigma-\sigma}=K_{2122}^{\sigma-\sigma}=\beta\delta$;
$K_{2221}^{\sigma-\sigma}=K_{2212}^{\sigma-\sigma}=\gamma\delta$.
The high-order correlators $K_{121122}^{\sigma-\sigma-\sigma}$ and
$K_{211122}^{\sigma-\sigma-\sigma}$ are exactly equal to zero as
they are the solution of the linear homogeneous system of equations
with zero initial conditions.

Let us discuss the changing of entanglement properties of initial
state during the system time evolution. A standard measure of the
entanglement is the concurrence $C$
\cite{Nizama},\cite{Wootters},\cite{Contreras-Pulido}. For each pure
state the entanglement $E$ is defined as the entropy of either of
the two subsystems $A$ and $B$ \cite{Bennett_1}:

\begin{eqnarray}
E(\psi)=-Tr(\rho_{A}log_{2}\rho_{A})=-Tr(\rho_{B}log_{2}\rho_{B}).\label{ent}
\end{eqnarray}

Here $\rho_{A}$ is the partial trace of $|\psi\rangle\langle\psi|$
over subsystem $B$, and $\rho_{B}$ has the similar meaning. The
entanglement of the mixed state $\rho$ is then defined as the
average entanglement of the pure states of the decomposition,
minimized over all decompositions of $\rho$:

\begin{eqnarray}
E(\rho)=min\sum_{i}p_{i}E(\psi_i).
\end{eqnarray}

To determine the concurrence for the system of two single-level QDs
with two electrons with opposite spins one has to distinguish two
different situations.

\subsection{Subsystems $A$ and $B$ are QDs (I)}
We now consider in detail the situation when subsystems $A$ and $B$
are QDs: the first QD - dot with energy level $\varepsilon_1$ and
the the second QD - dot with energy level $\varepsilon_2$, directly
coupled to the reservoir (subsystem $C$) correspondingly.
Interaction between the second QD and the reservoir is switched on
at the initial time moment $t=0$. For the each dot four electronic
states are possible: $|0\rangle_l$, $|\uparrow\rangle_l$,
$|\downarrow\rangle_l$ and $|\uparrow\downarrow\rangle_l$, where
$l=1,2$. We are going to analyze entanglement between the electronic
states in subsystems $A$ and $B$ (the first and the second QD). The
concurrence for pure state $|\psi\rangle$ is determined as
$C_{I}=|\langle\psi|\widetilde{\psi}\rangle|$, where
$|\widetilde{\psi}\rangle$ is the "spin flipped" state
$|\psi\rangle$. For mixed state concurrence is
$C_{I}=max\{0,\lambda_1-\lambda_2-\lambda_3-\lambda_4\}$, where
$\{\lambda_m\}$ - square roots of eigenvalues of matrix
$\widetilde{\rho}\rho$ ($\widetilde{\rho}$ is the "spin flipped"
matrix $\rho$) arranged in the decreasing order. For initial
two-electron entangled pure state $|\psi\rangle$ [see
Ex.(\ref{wave})] with opposite spins one can define $C_{I}$

\begin{eqnarray}
C_{I}=|\alpha^{2}+\delta^{2}+2\beta\gamma|.
\label{con}\end{eqnarray}

We'll demonstrate that for arbitrary mixed state concurrence $C_{I}$
can be determined through the mean value of pair correlators
$K_{lrl^{'}r^{'}}^{\sigma-\sigma}$ particular combination

\begin{eqnarray}
C_{I}=\langle
K_{1111}^{\sigma-\sigma}+K_{1221}^{\sigma-\sigma}+K_{2112}^{\sigma-\sigma}+K_{2222}^{\sigma-\sigma}\rangle.
\end{eqnarray}

Let us introduce operator $\widehat{K}^{'}$, which can be expressed
in terms of pair correlations operators:
\begin{eqnarray}
\widehat{K}^{'}=\widehat{K}_{1111}^{\sigma-\sigma}+\widehat{K}_{1221}^{\sigma-\sigma}+\widehat{K}_{2112}^{\sigma-\sigma}+\widehat{K}_{2222}^{\sigma-\sigma}.
\end{eqnarray}

Acting by operator $\widehat{K}^{'}$ on the wave function
$|\psi\rangle$ [see Ex.(\ref{wave})] one obtain "spin flipped" wave
function $|\widetilde{\psi}\rangle$

\begin{eqnarray}
\widehat{K}^{'}|\psi\rangle=|\widetilde{\psi}\rangle.
\end{eqnarray}

For any wave function $|\psi\rangle$:

\begin{eqnarray}
\langle\psi|\widehat{K}^{'}|\psi\rangle=\langle\psi|\widetilde{\psi}\rangle=C_{I}.
\end{eqnarray}

One can also find wave functions $\psi_{i} (i=1,2,3,4)$ for
two-electron states with opposite spins, which are the eigenstates
of the Hamiltonian $\hat{H}_{D}$
\begin{eqnarray}
|\psi_{i}\rangle=\alpha_{i}|\uparrow\downarrow\rangle|0\rangle+\beta_{i}|\downarrow\rangle|\uparrow\rangle+
\gamma_{i}|\uparrow\rangle|\downarrow\rangle+\delta_{i}|0\rangle|\uparrow\downarrow\rangle.
\label{wai}\end{eqnarray}

The corresponding eigenvalues $E_{i}$ can be determined from
equation

\begin{eqnarray}
det(\widehat{H}-E\widehat{I})=0,
\end{eqnarray}

where $\widehat{I}$ is the unity matrix and

\begin{eqnarray}
\widehat{H}= \left(\begin{array}{ccccc}
2\varepsilon_1+U_{1} & T & -T & 0\\
T & \varepsilon_1+\varepsilon_2 & 0 & -T\\
-T & 0 & \varepsilon_1+\varepsilon_2 & T\\
0 & -T & T & 2\varepsilon_2+U_{2}\\
\end{array}\right).
\end{eqnarray}

In the case of resonant tunneling between the similar QDs
($\varepsilon_1=\varepsilon_2=\varepsilon_0$; $U_{1}=U_{2}=U$) the
coefficients $\alpha_{i}$, $\beta_{i}$, $\gamma_{i}$ and
$\delta_{i}$ for the ground singlet state can be obtained
analytically:

\begin{eqnarray}
\alpha&=&\delta=\frac{\sqrt{2}T}{\sqrt{4T^{2}+b^{2}}},\nonumber\\
\beta&=&\gamma=\frac{b}{\sqrt{2}\sqrt{4T^{2}+b^{2}}},
\end{eqnarray}

where

\begin{eqnarray}
b=\frac{U}{2}+\sqrt{\frac{U^{2}}{4}+4T^{2}}.
\end{eqnarray}

The energy of the ground state has the value
$E_{G}=\varepsilon_0+\frac{U}{2}-\sqrt{\frac{U^{2}}{4}+4T^{2}}$.

If $\{|\psi_i\rangle\}$ are the two-particle eigenfunctions for
electrons with opposite spins of the Hamiltonian $\widehat{H}$, two
particle density matrix can be expressed as
$\rho=\sum_{i}|\psi_i\rangle\langle\psi_i|p_{i}$. The following
relations take place: $\langle
\psi_j|\widehat{K}^{'}|\widetilde{\psi_i}\rangle=\delta_{ij}$ and
$\langle
\psi_{i^{'}}|\widehat{K}^{'2}|\psi_i\rangle=\delta_{ii^{'}}=\sum_{j}\langle
\psi_{i^{'}}|\widehat{K}^{'}|\widetilde{\psi_j}\rangle\langle\widetilde{\psi_j}|\widehat{K}^{'}|\psi_i\rangle$.

Let us prove that
\begin{eqnarray}
\langle \psi_j|\widetilde{\rho}\rho|\psi_i\rangle=\langle
\psi_j|\widehat{K}^{'}\rho\widehat{K}^{'}\rho|\psi_i\rangle.
\label{eq0}\end{eqnarray}

Really:

\begin{eqnarray}
\langle
\psi_j|(\widehat{K}^{'}\rho)^{2}|\psi_i\rangle&=&\sum_{i_{1}}\langle
\psi_j|\widehat{K}^{'}|\psi_{i_{1}}\rangle\langle
\psi_{i_{1}}|\widehat{K}^{'}|\psi_i\rangle
p_{i}p_{i_{1}}=\nonumber\\&=&\sum_{i_{1}}\langle
\psi_j|\widetilde{\psi}_{i_{1}}\rangle\langle
\psi_{i_{1}}|\widetilde{\psi}_i\rangle p_{i}p_{i_{1}}
\label{eq1}\end{eqnarray}

and

\begin{eqnarray}
\langle \psi_j|\widetilde{\rho}\rho|\psi_i\rangle=p_{i}\langle
\psi_j|\widetilde{\rho}|\psi_i\rangle=\sum_{i_{1}}p_{i}p_{i_{1}}\langle
\psi_j|\widetilde{\psi}_{i_{1}}\rangle\langle\widetilde{\psi}_{i_{1}}|\psi_i\rangle
\label{eq2}.\end{eqnarray}

Taking into account expression (\ref{wai}) and comparing expressions
(\ref{eq1}) and (\ref{eq2}), one can find that statement (\ref{eq0})
is valid. If $\widetilde{\lambda}_{p}$ are the eigenvalues of matrix
$\|\widetilde{\rho}\rho\|_{ji}$ and $\lambda_{m}$ are the
eigenvalues of matrix $\|\widehat{K}^{'}\rho\|_{ji^{'}}$, then
$\lambda^{2}_{m}=\widetilde{\lambda}_{p}$ and
$\lambda_{m}=\pm\sqrt{\widetilde{\lambda}_{p}}$. So,

\begin{eqnarray}
Tr(\widehat{K}^{'}\rho)=\sum_{m}\lambda_{m}.
\end{eqnarray}

The concurrence $C_{I}$ \cite{Wootters} is expressed through
$\widetilde{\lambda}_{p}$, arranged in decreasing order, as
$C_{I}=max\{0,\sqrt{\widetilde{\lambda}_{1}}-\sqrt{\widetilde{\lambda}_{2}}-\sqrt{\widetilde{\lambda}_{3}}-\sqrt{\widetilde{\lambda}_{4}}\}$.
Finally

\begin{eqnarray}
C_{I}=max\{0,Tr(\widehat{K}^{'}\rho)\}=max\{0,\langle\widehat{K}^{'}\rangle\}.
\label{eq5}\end{eqnarray}

We would like to point out that for a pure state $|\psi\rangle$ [see
Ex.(\ref{wave})] the entanglement of subsystem $A$ (first QD) with
the surrounding subsystems $B$ (second QD) and $C$ (reservoir) can
be expressed as:

\begin{eqnarray}
E(\rho_{A})=Tr_{BC}\rho_{ABC}=-\alpha^{2}log_{2}\alpha^{2}-\nonumber\\-\beta^{2}log_{2}\beta^{2}-\gamma^{2}log_{2}\gamma^{2}-\delta^{2}log_{2}\delta^{2}
\end{eqnarray}

In the absence of Coulomb interaction ($U=0$) for symmetric QDs in
the singlet two-electron state one can find that
$\alpha=\delta=\beta=\gamma=1/2$. In this case the concurrence
$C_{I}$ [see Ex.(\ref{con})] is equal to zero, but entanglement
$E(\rho_{A})=2$. This means that subsystems $A$ and $B$ (first and
second QDs) are disentangled, but electrons in the first QD are
maximally entangled with the reservoir (subsystem $C$).

\subsection{Subsystems $A$ and $B$ are opposite spin systems (II)}

Let us now consider the other situation when the subsystem $A$
corresponds to the electrons with spin projections $S_{Z}=+\sigma$
and subsystem $B$ - to electrons with spin projections
$S_{Z}=-\sigma$. Each particle with particular spin projection can
be found in the first or in the second QD. If particle is found in
the first QDs we attribute $+1$ to this state and if it is situated
in the second QD we attribute $-1$. This case is similar to the
problem of two interacting "frozen spins" at neighboring sites. For
example, the states $|\uparrow\rangle|\uparrow\rangle$ and
$|\downarrow\rangle\downarrow\rangle$ in the "frozen spin" problem
corresponds to the states $|\uparrow\downarrow\rangle|0\rangle$ and
$|0\rangle|\downarrow\uparrow\rangle$ in the considered situation.
The "spin flip" in the "frozen spin" problem corresponds to
permutation of QDs. The pure state $|\psi\rangle$ [see
Ex.(\ref{wave})] is then transformed to the state
$|\widetilde{\widetilde{\psi}}\rangle$

\begin{eqnarray}
|\widetilde{\widetilde{\psi}}\rangle=\alpha|0\rangle|\downarrow\uparrow\rangle-\beta|\downarrow\rangle|\uparrow\rangle-\gamma|\uparrow\rangle|\downarrow\rangle+\delta|\uparrow\downarrow\rangle|0\rangle.
\end{eqnarray}

The concurrence $C_{II}$ in the pure state can be determined as
usual $C_{II}=|\langle\psi|\widetilde{\widetilde{\psi}}\rangle|$.
For the state $|\psi\rangle$ [see Ex.(\ref{wave})] the concurrence
$C_{II}=|2\alpha\delta-2\beta\gamma|$.

Let us introduce operator $\widehat{K}^{''}$, which can be expressed
in terms of pair correlations operators
\begin{eqnarray}
\widehat{K}^{''}=\widehat{K}_{1212}^{\sigma-\sigma}+\widehat{K}_{2121}^{\sigma-\sigma}-\widehat{K}_{1221}^{\sigma-\sigma}-\widehat{K}_{2112}^{\sigma-\sigma}.
\end{eqnarray}

One can obtain the "spin flipped" state wave function
$|\widetilde{\widetilde{\psi}}\rangle$

\begin{eqnarray}
\widehat{K}^{''}|\psi\rangle=|\widetilde{\widetilde{\psi}}\rangle.
\end{eqnarray}

The concurrence $C_{II}$ for the pure state is the mean value of
operator $\widehat{K}^{''}$

\begin{eqnarray}
\langle\psi|\widehat{K}^{''}|\psi\rangle=\langle\psi|\widetilde{\widetilde{\psi}}\rangle=C_{II}.
\end{eqnarray}

Similarly to the previous case the following relations are valid:
$\langle
\psi_j|\widehat{K}^{''}|\widetilde{\widetilde{\psi_i}}\rangle=\delta_{ij}$;
$\langle
\psi_{i^{'}}|\widehat{K}^{''2}|\psi_i\rangle=\delta_{ii^{'}}$ and

\begin{eqnarray}
\langle
\psi_j|\widetilde{\widetilde{\rho}}\rho|\psi_i\rangle=\langle
\psi_j|\widehat{K}^{''}\rho\widehat{K}^{''}\rho|\psi_i\rangle.
\label{eq4}\end{eqnarray}

The concurrence $C$ \cite{Wootters} expressed through
$\widetilde{\widetilde{\lambda}}_{p}$ (eigenvalues of matrix
$\|\widetilde{\widetilde{\rho}}\rho\|_{ji}$), arranged in decreasing
order, is
$C_{II}=max\{0,\sqrt{\widetilde{\widetilde{\lambda}}_{1}}-\sqrt{\widetilde{\widetilde{\lambda}}_{2}}-\sqrt{\widetilde{\widetilde{\lambda}}_{3}}-\sqrt{\widetilde{\widetilde{\lambda}}_{4}}\}$.
The expression (\ref{eq4}) allows to determine the concurrence
$C_{II}$ through the average value of operator $\widehat{K}^{''}$

\begin{eqnarray}
C_{II}=max\{0,\langle\widehat{K}^{''}\rangle\}.
\end{eqnarray}

This definition is similar to the obtained expression (\ref{eq5}).
Consequently, the entanglement [see Ex.(\ref{ent})] for the pure
state is:

\begin{eqnarray}
E(\rho_{A})=-\eta_{1}log_{2}\eta_{1}-\eta_{2}log_{2}\eta_{2},
\end{eqnarray}

where $\eta_i$ are the eigenvalues of matrix
$\|\rho_{\sigma}\|_{ij}$:

\begin{eqnarray}
\rho_{\sigma}= \left(\begin{array}{ccccc}
\alpha^{2}+\beta^{2} & \beta\delta+\alpha\gamma\\
\beta\delta+\alpha\gamma & \gamma^{2}+\delta^{2}\\
\end{array}\right)
\end{eqnarray}

with eigenvalues
$\eta_{1,2}=\frac{1}{2}\pm\frac{1}{2}\sqrt{1-C_{II}^{2}}$. In this
case the entanglement is always equal to zero for $C_{II}=0$. The
subsystems with particles with opposite spins are completely
disentangled for a pure singlet state in symmetric QDs in the
absence of Coulomb interaction.

Let us also introduce the dynamical concurrence $C_{I}(t)$ and
$C_{II}(t)$ for both considered cases in terms of pair correlation
functions:

\begin{eqnarray}
C_{I}(t)=max\{0,\langle\widehat{K}^{'}(t)\rangle\}
\end{eqnarray}
and

\begin{eqnarray}
C_{II}(t)=max\{0,\langle\widehat{K}^{''}(t)\rangle\}.
\end{eqnarray}

In this paper we are interested in the initially entangled pure
singlet electron state time evolution in the system of coupled QDs
with Coulomb correlations, which is the system ground state.
Interaction with the reservoir is switched on at the $t=0$. Our
model corresponds to the experimental situation when coupling
between vertically aligned strongly interacting QDs and substrate is
extremely weak in comparison with the coupling strength with another
lead (for example STM tip).

The obtained results for singlet entangled state relaxation are
discussed in the next section.

\section{Results and discussion}

The behavior of filling numbers time evolution depends on the
initial conditions, which are directly determined by the parameters
of the system: energy levels spacing, the Coulomb interaction and
interdots coupling values. We also analyzed the concurrence and pair
correlation functions time evolution.

\begin{figure}[h]
\includegraphics[width=80mm]{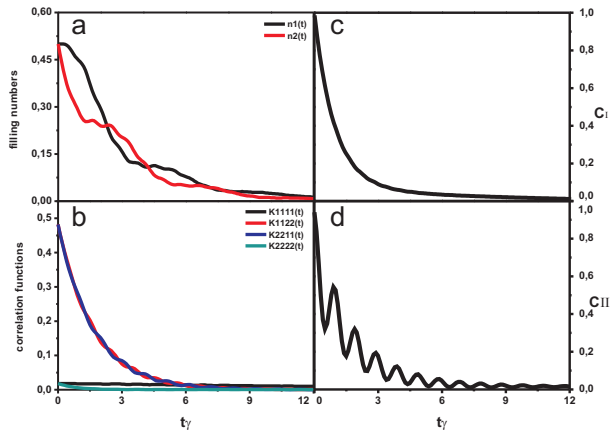}%
\caption{(Color online) a). Entangled state filling numbers time
evolution; b). Entangled state correlation functions time evolution;
c). Concurrence $C_{I}$ time evolution; d). Concurrence $C_{II}$
time evolution. The parameters values:
$\varepsilon_1=\varepsilon_2=2.0$; $U_1=U_2=6.0$; $T=0.6$;
$\gamma=0.3$. } \label{Fig.2}
\end{figure}

We start by discussing the resonant case:
$\varepsilon_1=\varepsilon_2$ (see Fig.\ref{Fig.2}). At the initial
time moment entangled electronic state demonstrates that charge is
equally distributed between the energy levels in the QDs
[$n_1(0)=n_2(0)$]. Correlation functions
$K_{1122}^{\sigma-\sigma}(0)$ and $K_{2211}^{\sigma-\sigma}(0)$ also
have the same values which strongly exceed the values of diagonal
correlation functions $K_{1111}^{\sigma-\sigma}(0)$ and
$K_{2222}^{\sigma-\sigma}(0)$ (see Fig.\ref{Fig.2}b). Such behavior
points on the magnetization of the system. For magnetic impurities
with definite value of spin projection $S_{z}=\pm1/2$, correlators
$K_{iiii}^{\sigma-\sigma}\ll1$ (turns to zero). From the other point
of view the behavior of pair correlators can be treated as the
Hund's rule for the electron filling numbers in coupled QDs with
Coulomb correlations. Electron filling numbers $n_{1}(t)$ and
$n_{2}(t)$ time evolution demonstrate multiple charge redistribution
between the QD's energy levels (see Fig.\ref{Fig.2}a).

The dynamical concurrences $C_{I}(t)$ and $C_{II}(t)$ are
demonstrated in the Fig.\ref{Fig.2}c,d. $C_{I}(t)$ reveals monotonic
decreasing during the relaxation processes. Dynamical concurrence
$C_{II}(t)$ demonstrates well pronounced oscillations which
amplitude decreases with time.

\begin{figure}[h]
\includegraphics[width=80mm]{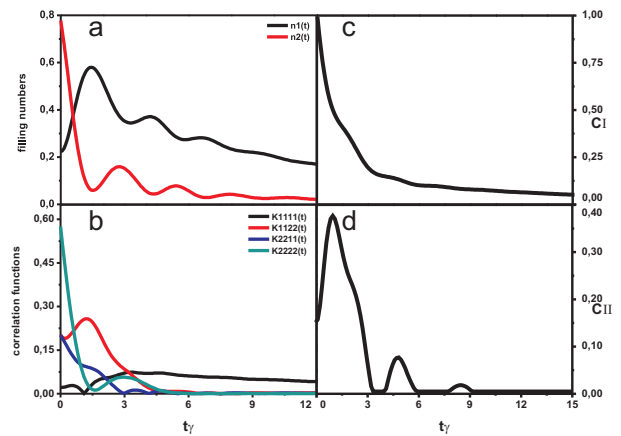}%
\caption{(Color online)  a). Entangled state filling numbers time
evolution; b). Entangled state correlation functions time evolution;
c). Concurrence $C_{I}$ time evolution; d). Concurrence $C_{II}$
time evolution. The parameters values: $\varepsilon_1=3.5$;
$\varepsilon_2=2.0$; $U_1=1.0$; $U_2=1.0$; $T=0.6$; $\gamma=0.3$.}
\label{Fig.3}
\end{figure}

\begin{figure}[h]
\includegraphics[width=80mm]{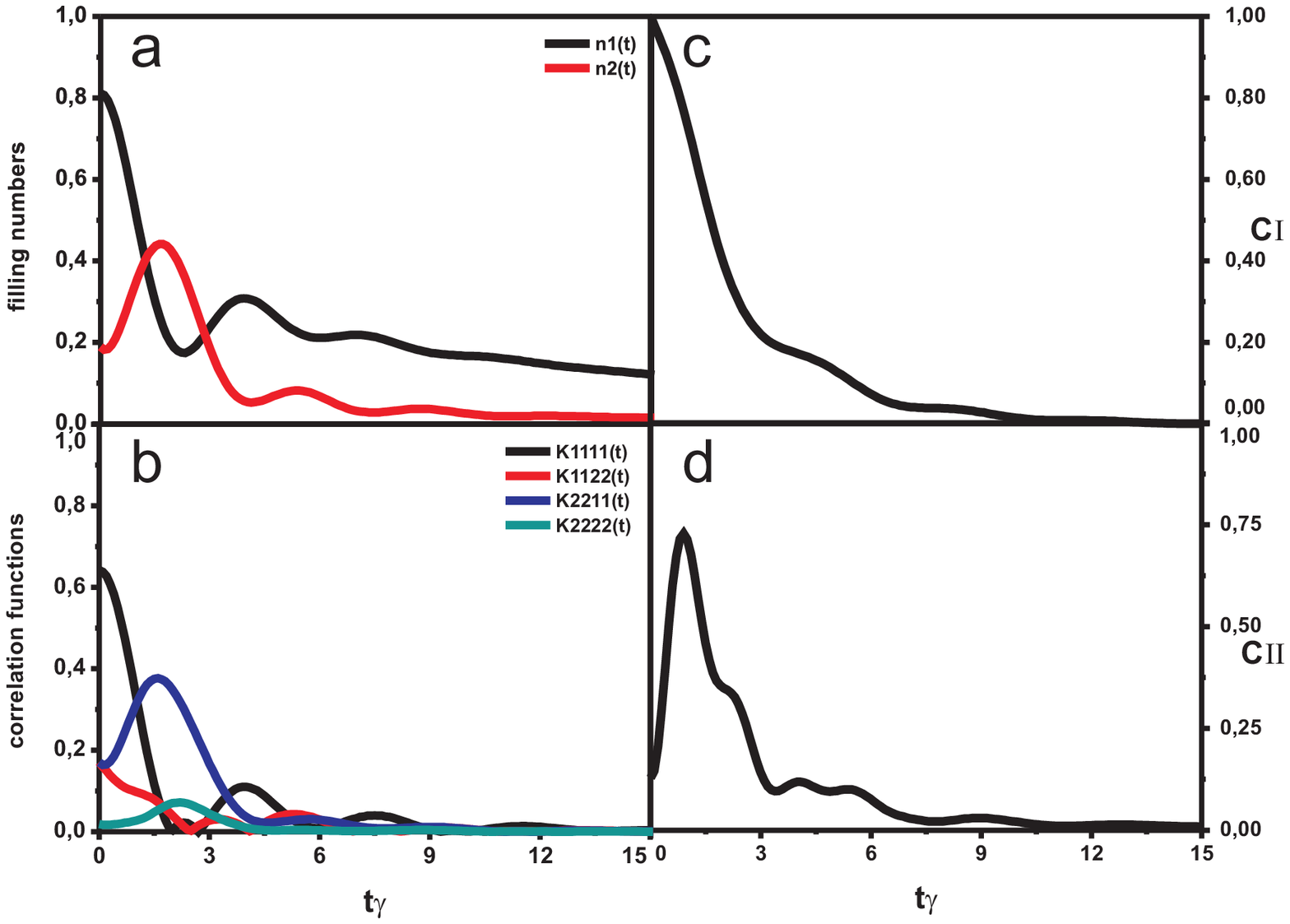}%
\caption{(Color online) a). Entangled state filling numbers time
evolution; b). Entangled state correlation functions time evolution;
c). Concurrence $C_{I}$ time evolution; d). Concurrence $C_{II}$
time evolution. The parameters values: $\varepsilon_1=2.0$;
$\varepsilon_2=3.5$; $U_1=0.75$; $U_2=0.75$; $T=0.6$; $\gamma=0.3$.}
\label{Fig.4}
\end{figure}

\begin{figure}[h]
\includegraphics[width=80mm]{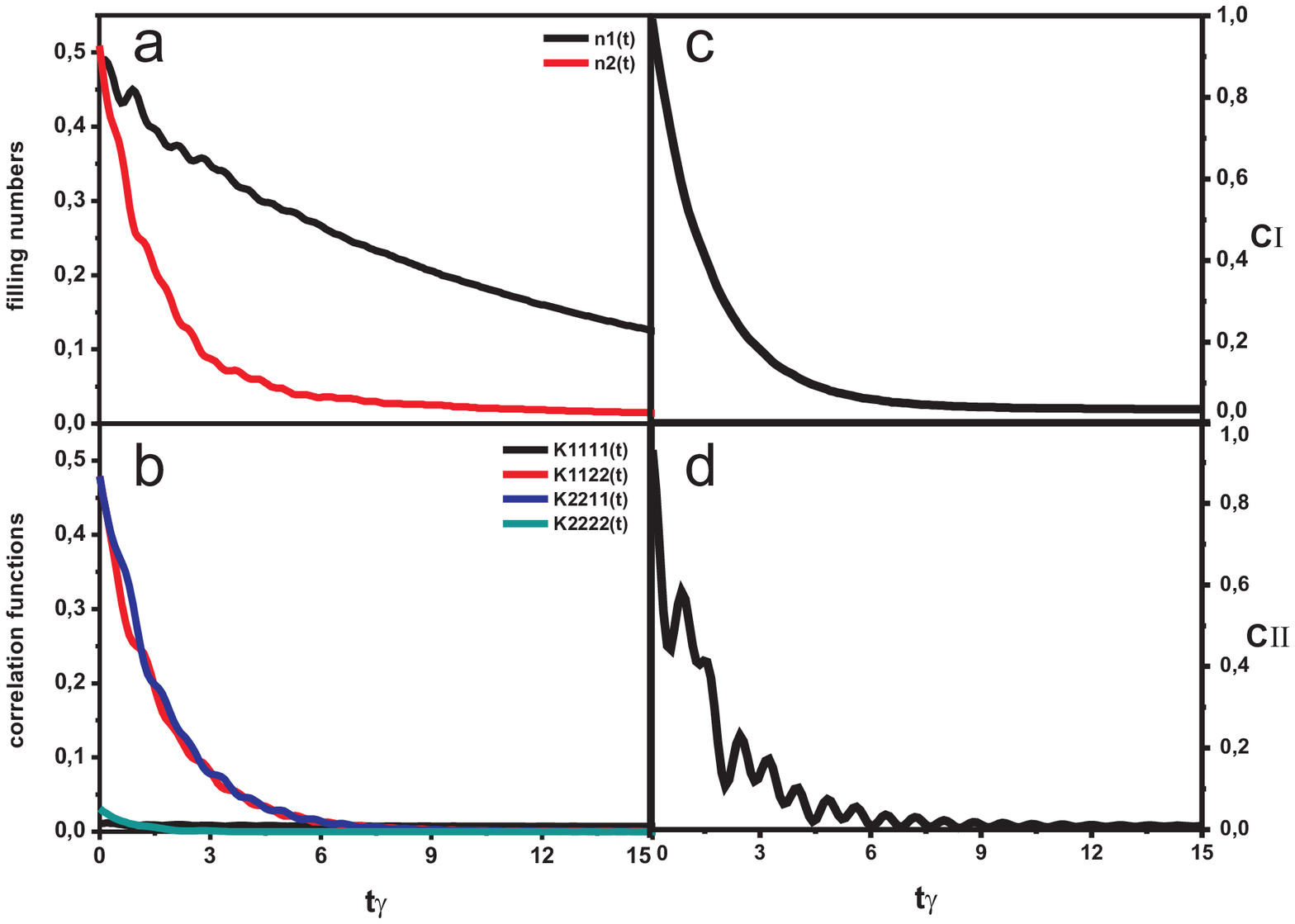}%
\caption{(Color online)  a). Entangled state filling numbers time
evolution; b). Entangled state correlation functions time evolution;
c). Concurrence $C_{I}$ time evolution; d). Concurrence $C_{II}$
time evolution. The parameters values: $\varepsilon_1=3.5$;
$\varepsilon_2=2.0$; $U_1=6.0$; $U_2=6.0$; $T=0.6$; $\gamma=0.3$.}
\label{Fig.5}
\end{figure}

\begin{figure}[h]
\includegraphics[width=60mm]{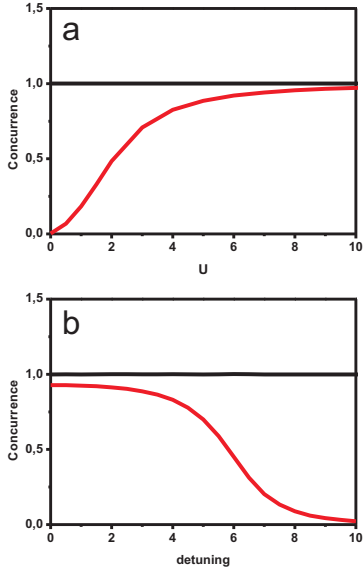}%
\caption{(Color online) (a)Concurrences $C_{I}(0)$ (black line) and
$C_{II}(0)$ (red line) as a functions of Coulomb interaction value
at $t=0$. Parameters values: $\varepsilon_1=3.5$;
$\varepsilon_2=2.0$; $T=0.6$; $\gamma=0.3$. (b) Concurrences
$C_{I}(0)$ (black line) and $C_{II}(0)$ (red line) as a functions of
energy levels detuning at $t=0$. The parameters values: $U_1=6.0$;
$U_2=6.0$; $T=0.6$; $\gamma=0.3$. } \label{Fig.6}
\end{figure}

Now we'll discuss the filling numbers time evolution both in the
case of the positive ($|\xi|=\varepsilon_1-\varepsilon_2>0$) (see
Fig.\ref{Fig.3}) and negative
($|\xi|=\varepsilon_1-\varepsilon_2<0$) (see Fig.\ref{Fig.4})
initial detunings between energy levels in the dots with rather weak
Coulomb interaction ($U<|\xi|$).

For the positive detuning at the initial time moment charge in the
system is mostly localized on the lower energy level
[$n_1(0)<n_2(0)$] and the value of correlation function
$K_{2222}^{\sigma-\sigma}(0)$ is the largest one in the system (see
Fig.\ref{Fig.3}a,b). Filling numbers amplitude $n_1(t)$ continue
being smaller than the filling numbers amplitude on the lower energy
level $n_2(t)$ during the system evolution until the amplitude of
correlation function $K_{2222}^{\sigma-\sigma}(0)$ exceeds
amplitudes of other pair correlation functions. During further time
evolution the appearance of dynamical inverse occupation occurs and
correlation functions $K_{1122}^{\sigma-\sigma}(t)$ and
$K_{2211}^{\sigma-\sigma}(t)$ exceed the values of diagonal
correlation functions $K_{1111}^{\sigma-\sigma}(t)$ and
$K_{2222}^{\sigma-\sigma}(t)$.

In the case of negative detuning at the initial time moment charge
in the system is mostly localized on the lower energy level
[$n_2(0)<n_1(0)$] and the value of correlation function
$K_{1111}^{\sigma-\sigma}(0)$ is the largest one in the system (see
Fig.\ref{Fig.4}a,b). Filling numbers amplitude $n_2(t)$ continue
being the smaller one during the system evolution until the
amplitude of correlation function $K_{1111}^{\sigma-\sigma}(0)$ is
the largest one. Relaxation reveals the appearance of dynamical
inverse occupation, which takes place when correlation function
$K_{2211}^{\sigma-\sigma}(t)$ exceeds the values of other
correlation functions. Further time evolution is governed by the
correlation function $K_{1111}^{\sigma-\sigma}(t)$ and demonstrates
the absence of dynamical inverse occupation.

The dynamical concurrences $C_{I}(t)$ and $C_{II}(t)$ are
demonstrated in the Fig.\ref{Fig.3}c,d and Fig.\ref{Fig.4}c,d.
$C_{II}(t)$ reveal non-monotonic behavior, while $C_{I}(t)$
monotonically decreases. For both signs of initial detuning
concurrence $C_{II}(t)$ rapidly reaches maximum value (formation of
a peak) at particular time moment when the electron density is
equally distributed between the QDs (see Fig.\ref{Fig.3}a and
Fig.\ref{Fig.4}a). For positive initial detuning $C_{II}(t)$ further
time evolution reveals non-monotonic behavior and $C_{II}(t)$ even
turns to zero at particular time intervals. This means that the
system switches between entangled and unentangled states. For
negative initial detuning $C_{II}(t)$ also demonstrates
non-monotonic behavior, but the system becomes unentangled only at
the final stage of charge relaxation.

Localized charge relaxation in the case of QDs with strong Coulomb
interaction ($U>|\xi|$) is depicted in the Fig.\ref{Fig.5}. At the
initial time moment entangled electronic state demonstrates that
charge is quite equally distributed between the energy levels in the
QDs [$n_1(0)\leq n_2(0)$]. Correlation functions
$K_{1122}^{\sigma-\sigma}(0)$ and $K_{2211}^{\sigma-\sigma}(0)$ have
the same values which strongly exceed the values of diagonal
correlation functions $K_{1111}^{\sigma-\sigma}(0)$ and
$K_{2222}^{\sigma-\sigma}(0)$ (see Fig.\ref{Fig.5}b). Electrons
filling numbers $n_{1}(t)$ and $n_{2}(t)$ time evolution
demonstrates the formation of dynamical inverse occupation during
the relaxation process except the initial time moment (see
Fig.\ref{Fig.5}a). Similarly to the case when resonant tunneling
between the QDs occurs, non-diagonal correlation functions
$K_{1122}^{\sigma-\sigma}(t)$ and $K_{2211}^{\sigma-\sigma}(t)$
dominate in the charge relaxation process (see Fig.\ref{Fig.5}a). As
we mentioned above this effect can be treated as the fulfillment of
the Hund's rule in coupled QDs with Coulomb correlations.

The dynamical concurrence $C_{I}(t)$ amplitude decreases
monotonically with the decreasing of localized charge amplitude in
the system. Time evolution of $C_{II}(t)$ demonstrates well
pronounced oscillations which amplitude decreases with time.

Let us now discuss the behavior of concurrence at the initial time
moment. For initial ground two-electron state in coupled QDs
$C_{I}(0)$ and $C_{II}(0)$. Concurrences $C_{I}(0)$ (black line) and
$C_{II}(0)$ (red line) as a functions of Coulomb interaction value
and detuning between energy levels in the QDs are depicted in the
Fig.\ref{Fig.6}a and Fig.\ref{Fig.6}b correspondingly. Concurrence
$C_{I}(0)$ is always equal to unity for all values of initial
detuning and Coulomb interaction (see Fig.\ref{Fig.6}a and
Fig.\ref{Fig.6}b black line). $C_{II}(0)$ (red line) in the absence
of Coulomb interaction is equal to zero, the system is in the pure
unentangled state. The increasing of Coulomb interaction value
results in the growth of the $C_{II}(0)$ ($C_{II}(0)\rightarrow1$
for $U\rightarrow\infty$) (see Fig.\ref{Fig.6}a red line).
Consequently, Coulomb correlations in the system for infinitely
large $U$ lead to the formation of the fully entangled Bell's
electronic state at the initial time moment for coupled QDs.
Fig.\ref{Fig.6}b (red line) demonstrates that the concurrence
$C_{II}(0)$ monotonically decreases with the detuning growth.  The
entanglement disappears when energy levels spacing strongly exceeds
the Coulomb interaction value.

\section{Conclusion}

We demonstrated that for arbitrary mixed state the concurrence can
be determined from the average value of particular combinations of
localized electrons pair correlation functions. We obtained the
closed system of equations for time evolution of the localized
electrons filling numbers and pair correlation functions which
exactly takes into account all order correlations for localized
electrons.

From time dependence of electrons filling numbers and pair
correlation functions one can follow the time evolution of
concurrence and entanglement during the relaxation process. We
analyzed different possible ways to divide the investigated system
into two entangled subsystems. We have found special regimes when
dynamical concurrence demonstrates non-monotonic behavior during the
time evolution.

We revealed the appearance of dynamical inverse occupation of the
QDs energy levels and demonstrated that for large values of Coulomb
interaction non-diagonal pair correlation functions always exceed
the diagonal ones. When on-site Coulomb repulsion is smaller than
the energy levels detuning, correlation function of two electrons
with opposite spins localized in the QD with the lower energy level
exceeds all other correlation functions until the dynamical inverse
occupation occurs due to the relaxation process.

We also analyzed the dependence of initial value of concurrence on
the system parameters: Coulomb correlations value and energy levels
detuning. We revealed that concurrence $C_{II}(0)$ for large Coulomb
interaction values ($U\rightarrow\infty$) is close to unit for the
finit value of detuning. Concurrence $C_{I}(0)$ is always equal to
unity for all values of initial detuning and Coulomb interaction. We
also demonstrated the validity of Hund's rule for the two coupled
single level QDs when Coulomb interaction value is larger than the
energy levels detuning. Our results open up further perspectives in
solid state quantum information based on the controllable dynamics
of the entangled electron states.

This work was  supported by RFBR grants and by the National Grant of
Ministry of science and education.


\pagebreak

\end{document}